\documentclass[epj]{webofc}
\usepackage[utf8]{inputenc}
\usepackage[varg]{txfonts}  
\usepackage{booktabs}
\usepackage{xcolor}
\definecolor{darkred}{rgb}{0.4,0.0,0.0}
\definecolor{darkgreen}{rgb}{0.0,0.4,0.0}
\definecolor{darkblue}{rgb}{0.0,0.0,0.4}
\usepackage[bookmarks,linktocpage,colorlinks,
    linkcolor = darkred,
    urlcolor  = darkblue,
    citecolor = darkgreen]{hyperref}
\usepackage{amsmath}
\usepackage{amssymb}
\usepackage{amsfonts}
\usepackage{subfigure}
\wocname{EPJ Web of Conferences}
\woctitle{Lattice2017}

\newcommand{\Tr}{{\rm Tr\,}}

\newcommand{\Ort}{{\rm O\,}}

\newcommand{\USp}{{\rm USp\,}}
\newcommand{\U}{{\rm U\,}}
\newcommand{\SU}{{\rm SU\,}}

\newcommand{\eins}{\leavevmode\hbox{\small1\kern-3.8pt\normalsize1}}
\newcommand{\be}{\begin{eqnarray}}
\newcommand{\ee}{\end{eqnarray}}
\begin{document}
%%%%%%%%%%%%%%%%%%%%%%%%%%%%%%%%%%%%%%%%%%%%%%%%%%%%%%%%%%%%%%%%%%%%%%%%%%%%%
%
\selectlanguage{english}
%----------------------------------------------------------------------------
\title{%
Global Symmetries of Naive and Staggered Fermions in Arbitrary Dimensions
}
%----------------------------------------------------------------------------
\author{%
\firstname{Mario} \lastname{Kieburg}\inst{1}\fnsep\thanks{Speaker, Acknowledges support by the grant AK35/2-1 "Products of Random Matrices" of the German research council (DFG), \email{mkieburg@physik.uni-bielefeld.de}} \and
\firstname{Tim R.}  \lastname{W\"urfel}\inst{1}\fnsep\thanks{\email{twuerfel@physik.uni-bielefeld.de}}
% etc.
}
%----------------------------------------------------------------------------
\institute{%
  Fakult\"at f\"ur Physik, Universit\"at Bielefeld, Postfach 100131, 33501 Bielefeld, Germany
}
%----------------------------------------------------------------------------
\abstract{%
It is well-known that staggered fermions do not necessarily satisfy the same global symmetries as the continuum theory. We analyze the mechanism behind this phenomenon for arbitrary dimension and gauge group representation. For this purpose we vary the number of lattice sites between even and odd parity in each single direction. Since the global symmetries are manifest in the lowest eigenvalues of the Dirac operator, the spectral statistics and also the symmetry breaking pattern will be affected. We analyze these effects and compare our predictions with Monte-Carlo simulations of naive Dirac operators in the strong coupling limit. This proceeding is a summary of our work \cite{Kieburg:2017rrk}.
}
%----------------------------------------------------------------------------
\maketitle
%----------------------------------------------------------------------------
\section{Introduction}\label{intro}
The global symmetries of QCD-Dirac operators determine the number and the properties of the lightest pseudo-scalar mesons. Thus it is tremendously important that the discretized theory yields the same global symmetries in the continuum limit. For staggered fermions this in not necessarily guaranteed as found in \cite{Bruckmann:2008xr},\cite{Damgaard:2001fg}, at least at a finite lattice spacing. The reason is that the symmetry breaking pattern is changing \cite{Bruckmann:2008xr}, \cite{Damgaard:2001fg}, \cite{Kieburg:2014eca}, \cite{Bialas:2010hb}. The kind of change depends on the choice of the gauge group representation and the space-time dimension.\\
It is well-known that the global symmetries of the Dirac operator are manifested in the statistical properties of its smallest eigenvalues \cite{Verbaarschot:1994qf},\cite{Shuryak:1992pi}. Since the 90s we know that those eigenvalues can be modeled with random matrix theory (RMT)  \cite{Verbaarschot:1994qf},\cite{Verbaarschot:1994ip}. Thus it is a perfect tool to check whether the symmetry analysis correctly predicts the symmetries of Dirac operators in lattice simulations.\\
In the continuum a symmetry analysis was already done in \cite{DeJonghe:2012rw} resulting in ten different symmetry breaking patterns, which correspond to the Altland-Zirnbauer tenfold classification of RMT \cite{Dyson1962}, \cite{Zirnbauer:1996zz},\cite{AltlandZirnbauer}. The question how the discretized theory can be connected to the continuum was recently analysed for two dimensions in \cite{Kieburg:2014eca}. A shift of symmetries according to the number of lattice directions with even parity was found there. In \cite{Kieburg:2017rrk} we extended this discussion to arbitrary dimension and gauge group representation.\\
 In Sec. \ref{sec2} we derive the symmetry classification table for arbitrary dimension and show how the symmetry breaking patterns change for naive fermions on the lattice. In Sec. \ref{sec3} we confirm our predictions by comparing lattice simulations with RMT results in the strong coupling limit of the naive fermions.

%----------------------------------------------------------------------------
\section{Lattice QCD in $d$-dimensions}\label{sec2}
The Hilbert space $\mathcal{H}$ of QCD on a cubic lattice in $d$ space-time dimensions consists of three parts. We have $\mathcal{H}=\mathbb{C}^{d_r} \otimes \widehat{V} \otimes \mathbb{C}^{\lfloor d/2 \rfloor}$, where the first part is the color space. The variable $d_r$ is the dimension of the representation of the gauge group. The second part describes the cubic lattice $\widehat{V}=\bigotimes_{j=1}^{d} \mathbb{C}^{L_j}$ with $L_j$ being the number of lattice sites in direction $j$. The third part is the spinor space, where the $\gamma$-matrices act on. The dimension of $\mathcal{H}$ is given by $2^{\lfloor d/2 \rfloor}d_r V$, with $V$ the space-time volume $V=\prod_{j=1}^{d} L_j$.\\
Gauge fields at lattice point $x$ in lattice direction $\mu$ are represented by $U_\mu (x)$ and the wave function at lattice site $x$ is $\vert \psi (x) \rangle$. A translation operator $T_\mu$ of a naive discretization in the direction $\mu$ can be introduced by its action on the wave function at a fixed lattice site $x=(x_1,\ldots,x_d)\in \widehat{V}$, namely
\begin{equation}\label{transop}
T_\mu \vert \psi (x) \rangle = (-1)^{\delta_{\mu d}\delta_{x_d L_d}}U_\mu (x) \vert \psi (x+e_\mu )\rangle {\rm .}
\end{equation}
The naive Dirac operator in $d$ space-time dimensions is then
\begin{equation}\label{LatDiracop}
D= \sum_{\mu = 1}^{d} (T_\mu - T^{\dagger}_\mu)\gamma_\mu {\rm .}
\end{equation}
We want to recall that the generalized $\gamma$-matrices generate a Clifford algebra. Moreover they are traceless and Hermitian:
\begin{equation}\label{Cliffordrelation}
[\gamma_\mu ,\gamma_\nu ]_+ = 2\delta_{\mu \nu }\eins_{2^{\lfloor d/2 \rfloor}} {\rm ,} \quad \Tr \gamma_\mu =0 \quad {\rm  and} \quad \gamma^{\dagger}_\mu = \gamma_\mu {\rm .}
\end{equation}
The notation $[\cdot,\cdot]_+$ denotes the anticommutator. We employ the Euclidean version of the $\gamma$-matrices, since we consider only lattice QCD.\\
The symmetry analysis goes along the same lines as in the continuum case \cite{DeJonghe:2012rw}. We start with some general concepts known for Clifford algebras.  There is a chiral basis for even $d$, because of the non-triviality (not proportional to the identity matrix) of the matrix
\begin{equation}\label{chiralsymm}
\gamma^{(5)}=i^{-d(d-1)/2}\gamma_1 \gamma_2 \cdots \gamma_d {\rm .}
\end{equation}
This matrix is the same as in the continuum theory. The phase $i^{-d(d-1)/2}$ ensures the Hermiticity of $\gamma^{(5)}$. The Dirac operator anticommutes with $\gamma^{(5)}$:
\begin{equation}\label{chiralsymmDcomm}
 [D,\gamma^{(5)}]_+ =0 \quad  {\rm and} \quad \gamma^{(5)\dagger}=\gamma^{(5)}{\rm .}
\end{equation}
For odd $d$ there is no such symmetry.\\
There is an anti-unitary symmetry for the $\gamma$-matrices in any dimension,
\begin{equation}\label{antiunisym}
[C, i^{d(d-1)/2}\gamma_\nu ]_- = 0 \quad {\rm and} \quad [C,i^{d(d-1)/2}\gamma^{(5)}]_-=0
\end{equation}
with $[\cdot,\cdot]_-$ the commutator. Furthermore, the operator $C=K\chi$ consists of the complex conjugation operator $K$ and a product of $\gamma$-matrices denoted by $\chi$. The explicit form of $\chi$ depends on the space-time dimension $d$. The phase in Eq. \eqref{antiunisym} encodes the fact that $C$ may commute or anticommute depending on the space-time dimension $d$. The square of $C$ is
\begin{equation}\label{Csquare}
C^2=(-1)^{(d+2)(d+1)d(d-1)/8}\eins_{2^{\lfloor d/2 \rfloor}}
\end{equation}
which is the origin of the Bott periodicity of Clifford algebras \cite{BOTT1970353},\cite{Bott}. The anti-unitary symmetry for the Clifford algebra must be combined with the anti-unitary symmetries for the translation operators $T_\mu$. By definition the translation operator depends on the gauge link variables $U_\mu$ and, hence, on the representation of the underlying gauge group. The gauge group $\SU (N_c)$ has two particular representations, the fundamental and the adjoint representation. Let us underline that, although, we only concentrate on these two kinds of gauge theories the following discussion can be extended to arbitrary gauge groups and representations, see \cite{Kieburg:2017rrk},\cite{DeJonghe:2012rw}.  For gauge fields in the adjoint representation we find for $N_c >1$
\begin{equation}\label{realrepT}
[K, U]_- = 0 \quad \forall U \in \SU_{\rm a} (N_c>1) \quad \Rightarrow \quad C_{\rm lat}=K\chi{\rm .}
\end{equation}
For the fundamental representation only with $N_c=2$ a special commutation relation for all gauge elements can be obtained:
\begin{equation}\label{quatrepT}
[K\tau_2, U]_- = 0 \quad \forall U \in \SU_{\rm f} (N_c=2) \quad \Rightarrow \quad C_{\rm lat}=K\tau_2 \chi{\rm .}
\end{equation}
We define $\xi=\eins_{N_c^2-1}$ for the adjoint representation and $\xi=\tau_2$ for the fundamental representation with $N_c=2$. The constructed charge conjugation operator $C_{\rm lat}=K\xi \chi$ commutes with the Dirac operator:
\begin{equation}\label{CDcomm}
[C_{\rm lat},i^{d(d-1)/2}D]_-=0 \quad {\rm and} \quad C_{\rm lat}^{2}=(-1)^{(d+2)(d+1)d(d-1)/8}{\rm sign}[(K\xi\chi)^2]\eins_{d_{\mathcal{H}}}{\rm .}
\end{equation}
The square of $C_{\rm lat}$ determines wether there is a real or a quaternionic basis. We call the representation real, if $C^2_{\rm lat}=+\eins$ and quaternion if $C^2_{\rm lat}=-\eins$. For $N_c> 2$ in the fundamental representation we do not have a charge conjugation operator. We call this a complex representation.\\
As in two dimensions \cite{Kieburg:2014eca}  additional symmetries may appear depending on the lattice directions with even number of lattice sites.  Suppose the number of lattice sites $L_\mu$ in direction $\mu$ to be even. We can define the operator
\begin{equation}\label{GammaOp}
\Gamma_\mu \vert \psi(x)\rangle = (-1)^{x_\mu}\vert \psi(x)\rangle
\end{equation}
which is diagonal with eigenvalues $\pm 1$ and acts only on the $\widehat{V}$-part of the Hilbert space $\mathcal{H}$. This artifical operator satisfies the following commutation relations with the translation operators $T_\mu$
\begin{equation}\label{GammaOpTmuOPcomm}
[\Gamma_\mu , T_\mu ]_+ = 0 \quad {\rm and} \quad [\Gamma_\mu , T_\nu ]_- \overset{\mu \neq \nu}{=}0 {\rm .}
\end{equation}
Combining $\Gamma_\mu$ with the $\gamma$-matrices as follows
\begin{equation}\label{Gamma5muOp}
\Gamma^{(5)}_\mu = \Gamma_\mu \gamma_\mu{\rm ,}
\end{equation}
where we do not sum over $\mu$, we have the anticommutation relation with the Dirac operator
\begin{equation}\label{Gamma5OpDOp}
[\Gamma^{(5)}_\mu,D]_+ = \sum_{\nu = 1}^{d} [\Gamma_\mu \gamma_\mu , (T_\nu - T_\nu^{\dagger})\gamma_\nu]_+=0 {\rm .}
\end{equation}
We find such an operator $\Gamma^{(5)}$ for each lattice direction $\mu$ with an even number of lattice sites. Suppose we have $N_{\rm ev}$ directions with an even partition of lattice sites and denoting $\Gamma^{(5)}_{{\rm ev}+1}=\gamma^{(5)}$  if $d$ is even, the matrices $\{\Gamma^{(5)}_j\}_{j=1,\ldots,N}$, with $N=N_{\rm ev}+[d+1]_2$, generate a Clifford algebra, too. They are also Hermitian and traceless, i.e. for $i,j=1,\ldots, N_{{\rm ev}+1}$
\begin{equation}\label{Gamma5Cliff}
[\Gamma^{(5)}_i,\Gamma^{(5)}_j]_+ = 2\delta_{ij}\eins_{d_{\mathcal{H}}} {\rm ,} \quad \Tr \Gamma^{(5)}_j = 0 \quad {\rm and} \quad (\Gamma^{(5)}_j)^{\dagger}=\Gamma^{(5)}_j {\rm .}
\end{equation}
The commutation relations of $\Gamma^{(5)}_j$ with the charge conjugation operator $C_{\rm lat}$ reads
\begin{equation}\label{COpGamma5Op}
[C_{\rm lat},i^{d(d-1)/2}\Gamma^{(5)}_j]_-=0 \quad {\rm for} \quad j=1,\ldots,N,
\end{equation}
which is inherited from the $\gamma$-matrices.\\
Let us analyze the effect of the additional symmetries \eqref{Gamma5OpDOp} on the Dirac operator. We can find a unitary matrix $U\in \U (d_{\mathcal{H}})$ to transform the $\Gamma^{(5)}_j$ for $j=1,\ldots,N$ to
\begin{equation}\label{Trafo}
U\Gamma^{(5)}_j U^{\dagger}=
\begin{cases}
\eins_{d_{\mathcal{H}}/2^{N/2}} \otimes \gamma_j^{\prime}{\rm ,} & {\rm for} \; N \; {\rm even,} \\
\gamma^{(5)}_{\rm red} \otimes \gamma_j^{\prime} {\rm ,} &{\rm for} \; N \; {\rm odd.}
\end{cases}
\end{equation}
The case $N$ odd follows from the fact that also the operator $i^{N(N-1)/2}\Gamma^{(5)}_1 \cdots \Gamma^{(5)}_{N}$ is unitary, Hermitian and traceless, while the product of all $\gamma^\prime_j$ is proportional to the identity. The basis transformation \eqref{Trafo} together with Schur's Lemma \cite{schur} and the commutation relations \eqref{Gamma5OpDOp}, \eqref{Gamma5Cliff} lead to a reduced Dirac operator, i.e.
\begin{equation}\label{redDirac}
UD U^{\dagger}=
\begin{cases}
D_{\rm red} \otimes \gamma^{\prime (5)}{\rm ,} & {\rm for} \; N \; {\rm even,} \\
D_{\rm red} \otimes \eins_{2^{(N-1)/2}} {\rm ,} &{\rm for} \; N \; {\rm odd.}
\end{cases}
\end{equation}
\\
As long as the gauge group representation is complex, i.e. there is no anti-unitary symmetry, the case of  even $N$ yields a reduced Dirac operator $D_{\rm red}$ of dimension $d_{\rm red}=d-N_{\rm ev}$ whose global symmetries coincide with the three dimensional Dirac operator from the continuum. In the case of odd $N$ we have the symmetry $[D_{\rm red},\gamma^{(5)}_{\rm red}]_+=0$ in dimension $d_{\rm red}$ which coincides with the even dimensional continuum theories.\\
In the case of real or quaternion gauge group representation we have to transform the anti-unitary operator $C_{\rm lat}$ as well, i.e.
\begin{equation}\label{redClat}
C^\prime_{\rm lat}=UC_{\rm lat} U^{\dagger}= \underbrace{K\xi^\prime}_{=C_{\rm red}} \otimes \chi^\prime{\rm .}
\end{equation}
We now consider Eq. \eqref{COpGamma5Op} after transforming  with $U$, which is 
\begin{equation}\label{TrafoCommCGamma5}
[C^\prime_{\rm lat} , i^{d(d-1)/2}U\Gamma_j^{(5)}U^\dagger]_-=0 {\rm .}
\end{equation}
 We use Eq. \eqref{Trafo} and \eqref{redClat} to obtain a commutation relation of $\chi^\prime$ and $\gamma_j^{\prime}$ depending on $d$ and $N$. Then, one can find a representation of $\chi^\prime$ in terms of a product of $\gamma_j^{\prime}$ matrices. Thus we find the sign of $(K\chi)^2$ from which we can derive
\begin{equation}\label{chi2}
C_{\rm lat}^2=C_{\rm red}^2 \times (K\chi^\prime)^2{\rm .}
\end{equation}
Together with Eq. \eqref{CDcomm} this gives us the square of $C_{\rm red}$, namely
\begin{equation}\label{Cred2}
C^2_{\rm red} = (-1)^{(d_{\rm red}+2)(d_{\rm red}+1)d_{\rm red}(d_{\rm red}-1)/8} \times \text{sign}[(K\xi)^2] \times
\begin{cases}
\eins_{d_{\mathcal{H}}/2^{N/2}}  {\rm ,} & {\rm for} \; N \; {\rm even,} \\
\eins_{d_{\mathcal{H}}/2^{(N-1)/2}} {\rm ,} &{\rm for} \; N \; {\rm odd.}
\end{cases}
\end{equation}
For more details, see \cite{Kieburg:2017rrk}.\\
The symmetries for $D_{\rm red}$ with $C_{\rm red}$ are obtained directly from \eqref{CDcomm}, \eqref{redDirac} and \eqref{redClat}. Collecting all possible combinations of symmetries we obtain a Bott-perodic table in terms of space-time dimension and gauge group representation. This table coincides with the table one finds for the continuum theory of $d-N_{\rm ev}$ dimensions for arbitrary gauge group representation. Hence the dimension has only to be shifted by the number of lattice directions with an even partition.\\
Following from the above discussion we can write the reduced Dirac operator as
\begin{equation}\label{reducedDiracOp}
D_{\rm red}=\sum_{\mu = 1}^{N_{\rm ev}} D_\mu^{\rm (red)} + \sum_{\mu = N_{\rm ev}+1}^{d} D_\mu^{\rm (red)} \gamma_\mu
\end{equation}
with new covariant derivatives $D_\mu^{\rm (red)}$
\begin{equation}\label{covderivred}
D_\mu^{\rm (red)} \vert \psi (x) \rangle = (-1)^{\sum_{j=1}^{\mu-1} x_j}((-1)^{\delta_{jd}\delta_{x_d L_d}}U_\mu (x) \vert \psi (x+e_\mu) \rangle - (-1)^{\delta_{jd}\delta_{x_d L_1}} U_\mu^\dagger (x) \vert \psi (x-e_\mu ) \rangle )
\end{equation}
for $\mu \leq N$ and
\begin{equation}\label{covderivred2}
D_\mu^{\rm (red)} \vert \psi (x) \rangle = (-1)^{\sum_{j=1}^{N} x_j}((-1)^{\delta_{jd}\delta_{x_d L_d}}U_\mu (x) \vert \psi (x+e_\mu) \rangle - (-1)^{\delta_{jd}\delta_{x_d L_1}} U_\mu^\dagger (x) \vert \psi (x-e_\mu ) \rangle )
\end{equation}
for $\mu > N$. More details can be found in \cite{Kieburg:2017rrk}. The reduced Dirac operator is maximally Kramer's degenerate and only chiral if $d-N_{\rm ev}$ is even. We rediscover the staggered Dirac operator in the case $d=N_{\rm ev}$.\\
We know that the number of flavors for naive fermions may be different from the number of flavors for staggered fermions \cite{susskind}. This phenomenon can be also seen in the general setting. We know that the degeneracy of the Dirac operator $D$ depends on the number of lattice directions with an even partition of lattice sites $N_{\rm ev}$. The reduced Dirac operator acts on a Hilbert space of dimension $d_{\mathcal{H}}/d_{\rm tri}$ with $d_{\rm tri}=2^{\lfloor N/2 \rfloor}$. Therefore the characteristic polynomial of the lattice Dirac operator with quark mass $m$ is
\begin{equation}\label{symmbreakpat}
\det (D+m\eins_{d_{\mathcal{H}}})=
\begin{cases}
\det (D_{\rm red} + m \eins_{d_{\mathcal{H}}/d_{\rm tri}})^{d_{\rm tri}/2} \times \det (-D_{\rm red} + m \eins_{d_{\mathcal{H}}/d_{\rm tri}})^{d_{\rm tri}/2} {\rm ,} & {\rm for} \; N \; {\rm even,}\\
\det (D_{\rm red} + m \eins_{d_{\mathcal{H}}/d_{\rm tri}})^{d_{\rm tri}} {\rm ,} & {\rm for} \; N \; {\rm odd.}
\end{cases}
\end{equation}
Thus the number of physical flavors is enhanced by $d_{\rm tri}$ and the symmetry breaking patterns are those of the continuum theory in $d-N_{\rm ev}$ dimensions with $d_{\rm tri}N_f$ flavors. Consequently we obtain Table \ref{table1}.

\begin{table}[t!]
  \small
  \centering
  \caption{$N_{\rm eff}=d_{\rm tri}N_f$ with $d_{\rm tri}=2^{\lfloor N/2 \rfloor}$ degeneracy and $N_f$ the number of flavors. The last row indicates that the four symmetry breaking patterns from $8m$ to $8m+3$ in the quaternion representation reappear in the real representation for $8m+4$ to $8m+7$ and vice versa. The complex representation shows a $mod$ $2$ behaviour, meaning that the two possible symmetry breaking patterns just reappear periodicly for $8m+4$ and higher.}
  \label{table1}
  \begin{tabular}{lllll}\toprule
$d-N_{\rm ev}$ & real repr. & complex repr. & quaternion repr.  \\\midrule
& $\U(2N_{\rm eff})$ & $\U(N_{\rm eff}) \times \U(N_{\rm eff})$ & $\U(2N_{\rm eff})$ \\
$8m$   & \quad $\downarrow$  &  \quad  $\downarrow$  & \quad  $\downarrow$ \\  
& $\USp(2N_{\rm eff})$ & $\U(N_{\rm eff})$ & $\Ort(2N_{\rm eff})$ \\
\hline
& $\Ort(2N_{\rm eff})$ &$\U(2N_{\rm eff})$  & $\USp(2N_{\rm eff})$ \\
$8m+1$   & \quad  $\downarrow$  & \quad   $\downarrow$  & \quad  $\downarrow$ \\  
& $\U(N_{\rm eff})$ & $\U(N_{\rm eff}) \times \U(N_{\rm eff})$ & $\U(N_{\rm eff})$ \\
\hline
& $\Ort(2N_{\rm eff}) \times \Ort(2N_{\rm eff})$ & $\U(N_{\rm eff}) \times \U(N_{\rm eff})$ & $\USp(2N_{\rm eff}) \times \USp(2N_{\rm eff})$ \\
$8m+2$   & \quad  $\downarrow$  & \quad   $\downarrow$  & \quad  $\downarrow$ \\  
& $\Ort(2N_{\rm eff})$ & $\U(N_{\rm eff})$ & $\USp(2N_{\rm eff})$ \\
\hline
& $\Ort(2N_{\rm eff})$ & $\U(2N_{\rm eff})$ & $\USp(4N_{\rm eff})$ \\
$8m+3$   & \quad  $\downarrow$  & \quad   $\downarrow$  & \quad  $\downarrow$ \\  
& $\Ort(N_{\rm eff}) \times \Ort(N_{\rm eff})$ & $\U(N_{\rm eff}) \times \U(N_{\rm eff})$ & $\USp(2N_{\rm eff}) \times \USp(2N_{\rm eff})$ \\
\hline \hline
$8m+4+l$  & quat. repr. for $8m+l$  & see $8m+l$   & real repr. for $8m+l$ \\\bottomrule
  \end{tabular}
\end{table}

Finally we want to discuss the possibility of zero modes of the naive lattice Dirac operator. We find that QCD with a complex gauge group representation will never yield a Dirac operator with zero modes. For real or quaternion representations the generic zero modes can only appear in $d=1$ and $d=2$ dimensions. The latter case was indeed found in \cite{Kieburg:2014eca}. For more details see \cite{Kieburg:2017rrk}. We conclude that for $d>2$ naive lattice Dirac operators never show generic zero modes regardless of the considered gauge group representation.

\section{Comparison of lattice QCD with RMT}\label{sec3}

The given symmetry breaking patterns in Table~\ref{table1} indicate that any $d$-dimensional lattice shows the same eigenvalue statistics as the continuum theory in $d-N_{\rm ev}$ dimensions. We use RMT to verify this prediction. For this purpose we compare Monte Carlo simulations of naive quenched QCD lattice Dirac operators with random matrix theory results. We consider the Dirac operators in the strong coupling limit, meaning the gauge group elements are directly drawn from the Haar measure. Every lattice direction contains $3$ or $4$ sites, to distinguish between the even and odd cases. We have simulated lattices in $3$, $4$ and $5$ dimensions for gauge groups $\SU_{\rm f}(N_c >2)$ (complex repr.), $\SU_{\rm f}(N_c = 2)$ (quaternion repr.) and $\SU_{\rm a}(N_c >1)$ (real repr.). The dimension $d=2$ was done in \cite{Kieburg:2014eca}. The number of configurations we generated for the three- and four-dimensional lattices is $10^5$, while for the five-dimensional lattices we have done $10^3-10^4$ configurations. Each symmetry breaking pattern in Table~\ref{table1} can be identified with one of the ten Gaussian RMT models given in the Altland and Zirnbauer classification \cite{Dyson1962}, \cite{Zirnbauer:1996zz},\cite{AltlandZirnbauer}.\\

\begin{figure}[t!]
  \centering
\includegraphics[width=0.5\textwidth]{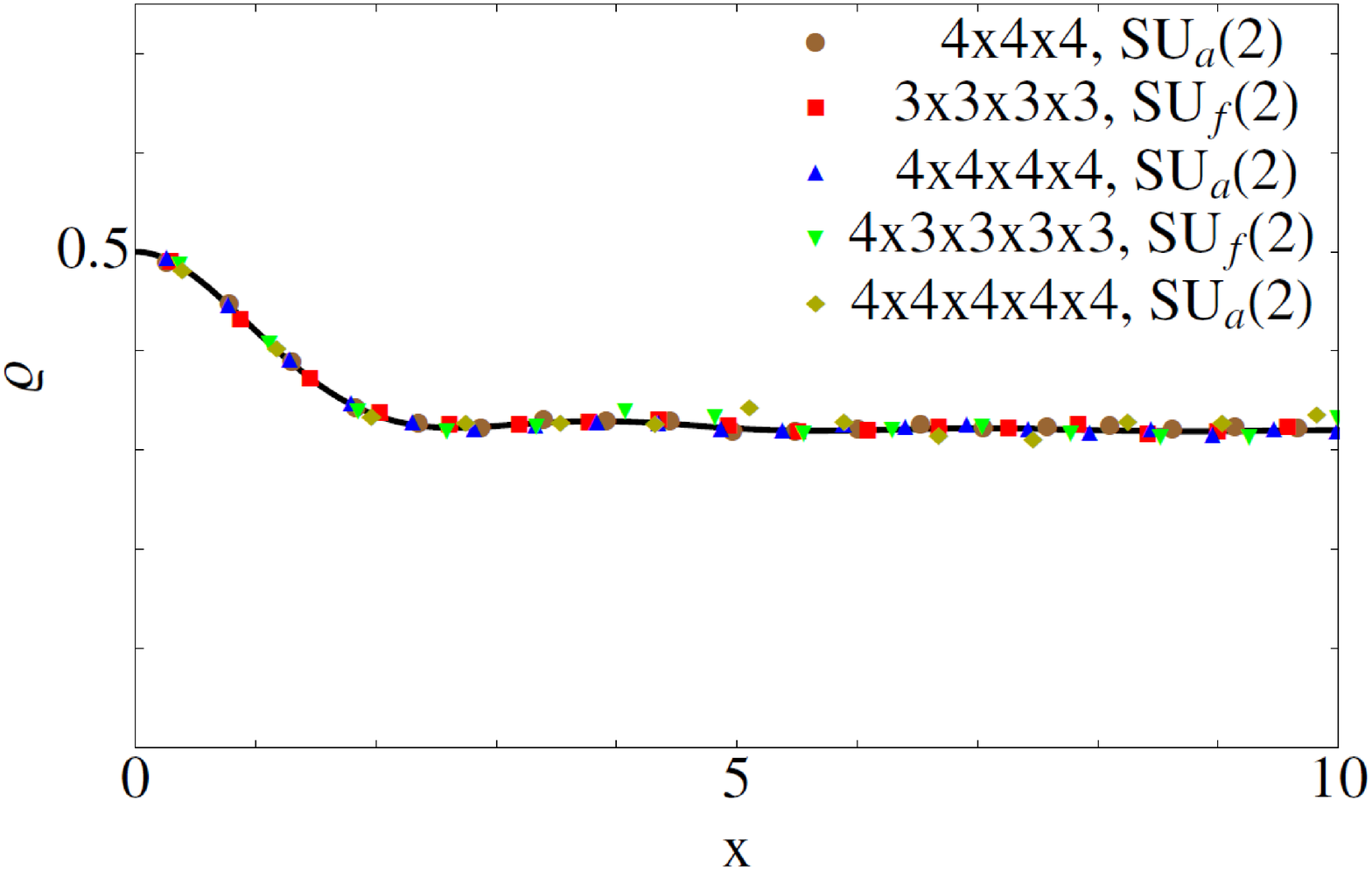}\includegraphics[width=0.5\textwidth]{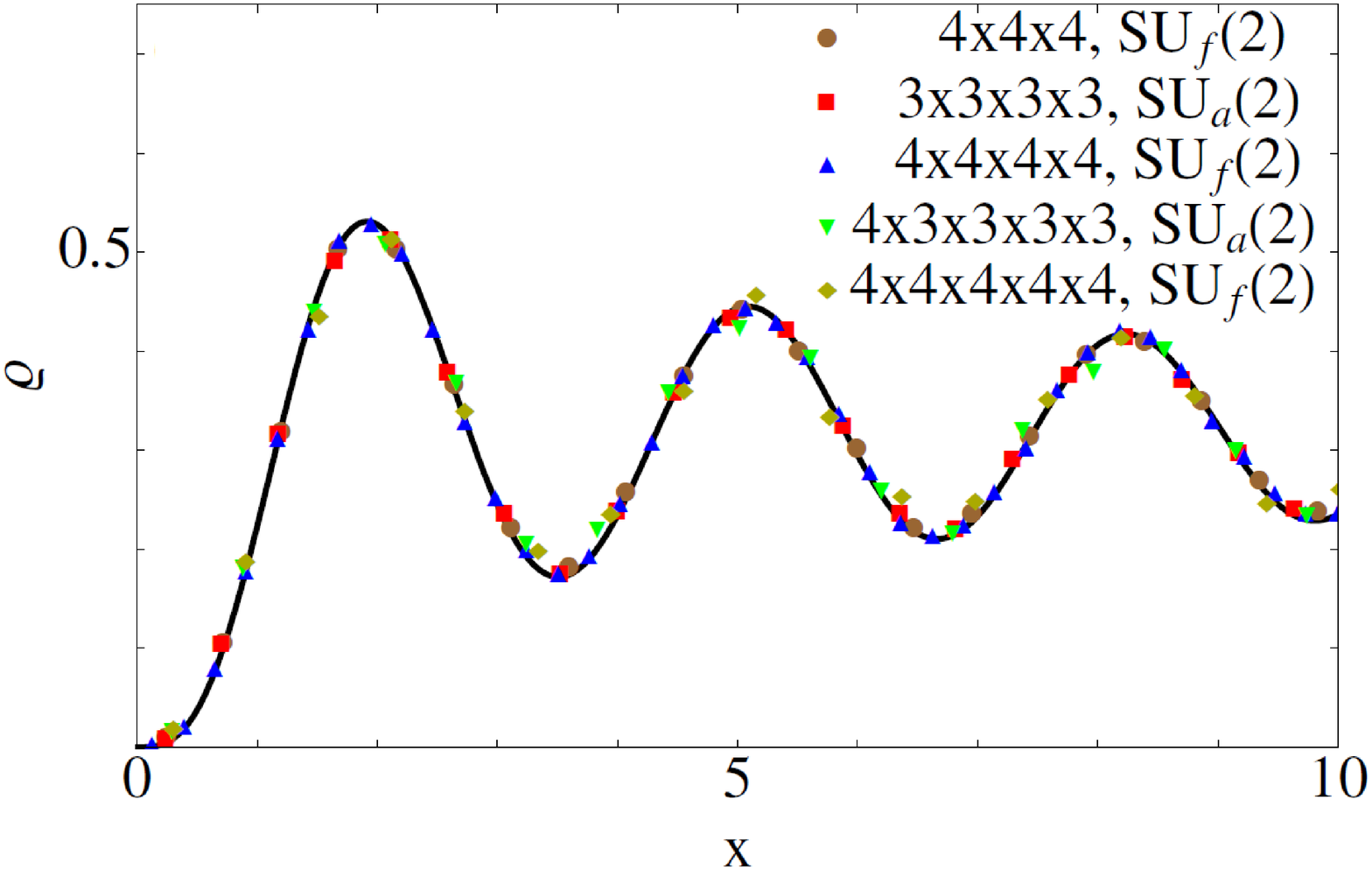}
  \caption{ Comparisons of some RMT predictions (black curves) with lattice simulations of three-, four- and five-dimensional quenched naive Dirac operators (colored symbols) in the strong coupling limit. The left figure corresponds to the Dyson index $\beta_D=1$ and the right figure corresponds to $\beta_D=4$. The abbreviations $\SU_f(N_c)$ and $\SU_a(N_c)$ stand for the fundamental and the adjoint representation of the gauge group $\SU(N_c)$, respectively. We show the microscopic level density $\rho^{(\beta_D)}_\nu(x)$ with $\beta_D=1,4$ as given in Eqs. \eqref{microdens} and \eqref{microdens2}. The simulations with staggered fermions are those when all numbers $L_\mu$ of lattice sites are even.}
\label{fig1}
\end{figure}

We employ two quantities known for these ten classes, namely the microscopic level density and the level spacing distribution. In Fig. \ref{fig1} we show only two classes out of the ten classes and concentrate on the microscopic level density. The other eight classes can be found in \cite{Kieburg:2017rrk}. The microscopic level densities in Fig. \ref{fig1} are results known from RMT \cite{Ivanov} and are given as
\begin{equation}\label{microdens}
\rho^{(1)}_\nu (x) = \frac{\vert x \vert}{2}(J_\nu^2(x) - J_{\nu +1}(x)J_{\nu -1}(x)) + \frac{1}{2}J_\nu (\vert x\vert) \left(1-\int^{\vert x \vert}_0 J_\nu (x^\prime)dx^\prime \right)
\end{equation}
for the Dyson index $\beta_D=1$ and
\begin{equation}\label{microdens2}
\rho^{(4)}_\nu (x) = \vert x \vert(J_{2\nu}^2(2x) - J_{2\nu +1}(2x)J_{2\nu -1}(2x)) -J_{2\nu} (2\vert x\vert) \left(\frac{1}{2}-\int_{\vert x \vert}^\infty J_{2\nu} (2x^\prime)dx^\prime \right)
\end{equation}
for the Dyson index $\beta_D=4$. We make use of the Bessel function of the first kind $J_\nu (x)$. The numerical data is fitted to the microscopic level density via a $\chi^2$-procedure.\\
The statistical error is smaller than $1\%$ for the three- and four-dimensional lattices and just a few procent for the five-dimensional ones due to the number of configurations we simulated. However there is a systematic error to consider: For computational reasons we had to choose the lattices sufficiently small, which leads to a small Thouless energy. But as we can see in Fig. \ref{fig1}, the Thouless energy must be larger than at least the first three eigenvalues. We recall that the Thouless energy represents the energy were the kinetic term in the physical system starts to show in the spectral statistics.

\section{Conclusions and Outlook}\label{conc}
We found a Bott-periodic classification of $d$-dimensional lattice QCD in the naive discretization. The classification holds for real, complex and quaternion representations of the gauge group $\SU(N_c)$ and matches with the continuum theory in $d-N_{\rm ev}$ dimensions, where $N_{\rm ev}$ denotes the number of lattice directions with even partition of lattice sites. We found an enhancement of flavors in the symmetry breaking patterns from $N_f$ to $d_{\rm tri}N_f$ with $d_{\rm tri}=2^{\lfloor N/2 \rfloor}$. The ten different symmetry classes appearing in the classification can be identified with the Altand-Zirnbauer tenfold way \cite{Dyson1962}, \cite{Zirnbauer:1996zz}, \cite{AltlandZirnbauer}. We compared the RMT models with Monte Carlo simulations for small lattices for all three gauge group representations and found very good agreement for the first few eigenvalues of the spectrum of the Dirac operator despite the small lattices. Furthermore we found that the Dirac operator has no exact zero modes in the naive and staggered discretization for any dimension $d>2$. Because of the Bott-periodicity staggered fermions have always the global symmetries of the continuum theory at $d=8$. An open question is how the global symmetries on the lattice change when the continuum limit is taken. It would be interesting to investigate if and how such a change is happening. Some work in this direction was done in \cite{Osborn:2003dr},\cite{Osborn:2010eq},\cite{Bialas:2010hb} and \cite{Bruckmann:2008xr}.

%%%%%%%%%%%%%%%%%%%%%%%%%%%%%%%%%%%%%%%%%%%%%%%%%%%%%%%%%%%%%%%%%%%%%%%%%%%%%
\end{document}